\documentclass[superscriptaddress,aps,prl,showkeys,showpacs,twocolumn,10pt,floatfix]{revtex4-1}
\usepackage[OT1,T1]{fontenc}
\usepackage{graphicx}
\usepackage{rotating}
\begin{document}

\title{Measurement of the $\vec{n}p \to d\pi^0\pi^0$ Reaction with Polarized
  Beam in the Region of
the $d^*(2380)$ Resonance}
\date{\today}

\newcommand*{\IKPUU}{Division of Nuclear Physics, Department of Physics and 
 Astronomy, Uppsala University, Box 516, 75120 Uppsala, Sweden}
\newcommand*{\ASWarsN}{Department of Nuclear Physics, National Centre for 
 Nuclear Research, ul.\ Hoza~69, 00-681, Warsaw, Poland}
\newcommand*{\IPJ}{Institute of Physics, Jagiellonian University, ul.\ 
 Reymonta~4, 30-059 Krak\'{o}w, Poland}
\newcommand*{\Edin}{School of Physics and Astronomy, University of Edinburgh, James Clerk Maxwell Building, Peter Guthrie Tait Road, Edinburgh EH9 3FD, UK}
\newcommand*{\PITue}{Physikalisches Institut, Eberhard--Karls--Universit\"at 
 T\"ubingen, Auf der Morgenstelle~14, 72076 T\"ubingen, Germany}
\newcommand*{\Kepler}{Kepler Center for Astro-- and Particle Physics, 
 University of T\"ubingen, Auf der Morgenstelle~14, 72076 T\"ubingen, Germany}
\newcommand*{\MS}{Institut f\"ur Kernphysik, Westf\"alische 
 Wilhelms--Universit\"at M\"unster, Wilhelm--Klemm--Str.~9, 48149 M\"unster, 
 Germany}
\newcommand*{\ASWarsH}{High Energy Physics Department, National Centre for 
 Nuclear Research, ul.\ Hoza~69, 00-681, Warsaw, Poland}
\newcommand*{\IITB}{Department of Physics, Indian Institute of Technology 
 Bombay, Powai, Mumbai--400076, Maharashtra, India}
\newcommand*{\Budker}{Budker Institute of Nuclear Physics of SB RAS, 
 11~akademika Lavrentieva prospect, Novosibirsk, 630090, Russia}
\newcommand*{\Novosib}{Novosibirsk State University, 2~Pirogova Str., 
 Novosibirsk, 630090, Russia}
\newcommand*{\IKPJ}{Institut f\"ur Kernphysik, Forschungszentrum J\"ulich, 
 52425 J\"ulich, Germany}
\newcommand*{\ZELJ}{Zentralinstitut f\"ur Engineering, Elektronik und 
 Analytik, Forschungszentrum J\"ulich, 52425 J\"ulich, Germany}
\newcommand*{\Erl}{Physikalisches Institut, 
 Friedrich--Alexander--Universit\"at Erlangen--N\"urnberg, 
 Erwin--Rommel-Str.~1, 91058 Erlangen, Germany}
\newcommand*{\ITEP}{Institute for Theoretical and Experimental Physics, State 
 Scientific Center of the Russian Federation, 25~Bolshaya Cheremushkinskaya, 
 Moscow, 117218, Russia}
\newcommand*{\Giess}{II.\ Physikalisches Institut, 
 Justus--Liebig--Universit\"at Gie{\ss}en, Heinrich--Buff--Ring~16, 35392 
 Giessen, Germany}
\newcommand*{\IITI}{Department of Physics, Indian Institute of Technology 
 Indore, Khandwa Road, Indore--452017, Madhya Pradesh, India}
\newcommand*{\HepGat}{High Energy Physics Division, Petersburg Nuclear Physics 
 Institute, 2~Orlova Rosha, Gatchina, Leningrad district, 188300, Russia}
\newcommand*{\HeJINR}{Veksler and Baldin Laboratory of High Energiy Physics, 
 Joint Institute for Nuclear Physics, 6~Joliot--Curie, 141980 Dubna, Moscow region, Russia}
\newcommand*{\Katow}{August Che{\l}kowski Institute of Physics, University of 
 Silesia, Uniwersytecka~4, 40-007, Katowice, Poland}
\newcommand*{\IFJ}{The Henryk Niewodnicza{\'n}ski Institute of Nuclear 
 Physics, Polish Academy of Sciences, 152~Radzikowskiego St, 31-342 
 Krak\'{o}w, Poland}
\newcommand*{\JLN}{Department of Physics, Malaviya National Institute of Technology Jaipur, JLN Marg, Jaipur--302017, Rajasthan, India}
\newcommand*{\JARA}{JARA--FAME, J\"ulich Aachen Research Alliance,
  Forschungszentrum J\"ulich, 52425 J\"ulich, and RWTH Aachen, 52056 Aachen,
  Germany}
\newcommand*{\INFN}{INFN, Laboratori 
  Nazionali di Frascati, Via E. Fermi, 40, 00044 Frascati (Roma), Italy}
\newcommand*{\Bochum}{Institut f\"ur Experimentalphysik I, Ruhr--Universit\"at 
 Bochum, Universit\"atsstr.~150, 44780 Bochum, Germany}
\newcommand*{\Tomsk}{Department of Physics, Tomsk State University, 36~Lenina 
 Avenue, Tomsk, 634050, Russia}
\newcommand*{\KEK}{High Energy Accelerator Research Organisation KEK, Tsukuba, 
 Ibaraki 305--0801, Japan} 
\newcommand*{\ASLodz}{Department of Astrophysics, National Centre for 
 Nuclear Research, Box 447, 90--950 {\L}\'{o}d\'{z}, Poland}

\author{P.~Adlarson}\altaffiliation[present address: ]{\Mainz}\affiliation{\IKPUU}
\author{W.~Augustyniak} \affiliation{\ASWarsN}
\author{W.~Bardan}      \affiliation{\IPJ}
\author{M.~Bashkanov\footnote{corresponding author\\ email address:
    mikhail.bashkanov@ed.ac.uk}} \altaffiliation[corresponding author:\\ ]{\EMAIL}  \affiliation{\Edin}\affiliation{\PITue}
\author{F.S.~Bergmann}  \affiliation{\MS}
\author{M.~Ber{\l}owski}\affiliation{\ASWarsH}
\author{H.~Bhatt}       \affiliation{\IITB}
\author{A.~Bondar}      \affiliation{\Budker}\affiliation{\Novosib}
\author{M.~B\"uscher}\altaffiliation[present address: ]{\PGI,\DUS}\affiliation{\IKPJ}
\author{H.~Cal\'{e}n}   \affiliation{\IKPUU}
\author{I.~Ciepa{\l}}   \affiliation{\IPJ}
\author{H.~Clement}     \affiliation{\PITue}\affiliation{\Kepler}
\author{E.~Czerwi{\'n}ski}\affiliation{\IPJ}
\author{K.~Demmich}     \affiliation{\MS}
\author{R.~Engels}      \affiliation{\IKPJ}
\author{A.~Erven}       \affiliation{\ZELJ}
\author{W.~Erven}       \affiliation{\ZELJ}
\author{W.~Eyrich}      \affiliation{\Erl}
\author{P.~Fedorets}  \affiliation{\IKPJ}\affiliation{\ITEP}
\author{K.~F\"ohl}      \affiliation{\Giess}
\author{K.~Fransson}    \affiliation{\IKPUU}
\author{F.~Goldenbaum}  \affiliation{\IKPJ}
\author{A.~Goswami}   \affiliation{\IITI}
\author{K.~Grigoryev}\altaffiliation[present address: ]{\Aachen}\affiliation{\IKPJ}\affiliation{\HepGat}
\author{C.--O.~Gullstr\"om}\affiliation{\IKPUU}
\author{L.~Heijkenskj\"old}\affiliation{\IKPUU}
\author{V.~Hejny}       \affiliation{\IKPJ}
\author{N.~H\"usken}    \affiliation{\MS}
\author{L.~Jarczyk}     \affiliation{\IPJ}
\author{T.~Johansson}   \affiliation{\IKPUU}
\author{B.~Kamys}       \affiliation{\IPJ}
\author{G.~Kemmerling}  \affiliation{\ZELJ}
\author{F.A.~Khan}      \affiliation{\IKPJ}
\author{G.~Khatri}       \affiliation{\IPJ}
\author{A.~Khoukaz}     \affiliation{\MS}
\author{D.A.~Kirillov}  \affiliation{\HeJINR}
\author{S.~Kistryn}     \affiliation{\IPJ}
\author{H.~Kleines}     \affiliation{\ZELJ}
\author{B.~K{\l}os}     \affiliation{\Katow}
\author{W.~Krzemie{\'n}}\affiliation{\IPJ}
\author{P.~Kulessa}     \affiliation{\IFJ}
\author{A.~Kup\'{s}\'{c}}\affiliation{\IKPUU}\affiliation{\ASWarsH}
\author{A.~Kuzmin}       \affiliation{\Budker}\affiliation{\Novosib}
\author{K.~Lalwani} \affiliation{\JLN}
\author{D.~Lersch}      \affiliation{\IKPJ}
\author{B.~Lorentz}     \affiliation{\IKPJ}
\author{A.~Magiera}     \affiliation{\IPJ}
\author{R.~Maier}       \affiliation{\IKPJ}\affiliation{\JARA}
\author{P.~Marciniewski}\affiliation{\IKPUU}
\author{B.~Maria{\'n}ski}\affiliation{\ASWarsN}
\author{H.--P.~Morsch}  \affiliation{\ASWarsN}
\author{P.~Moskal}      \affiliation{\IPJ}
\author{H.~Ohm}          \affiliation{\IKPJ}
\author{E.~Perez del Rio}\altaffiliation[present address: ]{\INFN}\affiliation{\PITue}
\author{N.M.~Piskunov}  \affiliation{\HeJINR}
\author{D.~Prasuhn}     \affiliation{\IKPJ}
\author{D.~Pszczel}     \affiliation{\IKPUU}\affiliation{\ASWarsH}
\author{K.~Pysz}        \affiliation{\IFJ}
\author{A.~Pyszniak}    \affiliation{\IKPUU}\affiliation{\IPJ}
\author{J.~Ritman}\affiliation{\IKPJ}\affiliation{\JARA}\affiliation{\Bochum}
\author{A.~Roy}         \affiliation{\IITI}
\author{Z.~Rudy}        \affiliation{\IPJ}
\author{O.~Rundel}      \affiliation{\IPJ}
\author{S.~Sawant}\affiliation{\IITB}\affiliation{\IKPJ}
\author{S.~Schadmand}   \affiliation{\IKPJ}
\author{I.~Sch\"atti--Ozerianska}\affiliation{\IPJ}
\author{T.~Sefzick}     \affiliation{\IKPJ}
\author{V.~Serdyuk} \affiliation{\IKPJ}
\author{B.~Shwartz}     \affiliation{\Budker}\affiliation{\Novosib}
\author{K.~Sitterberg}      \affiliation{\MS}
\author{R:~Siudak}       \affiliation{\IFJ}
\author{T.~Skorodko}\affiliation{\PITue}\affiliation{\Kepler}\affiliation{\Tomsk}
\author{M.~Skurzok}     \affiliation{\IPJ}
\author{J.~Smyrski}     \affiliation{\IPJ}
\author{V.~Sopov}       \affiliation{\ITEP}
\author{R.~Stassen}     \affiliation{\IKPJ}
\author{J.~Stepaniak}   \affiliation{\ASWarsH}
\author{E.~Stephan}     \affiliation{\Katow}
\author{G.~Sterzenbach} \affiliation{\IKPJ}
\author{H.~Stockhorst}  \affiliation{\IKPJ}
\author{H.~Str\"oher}   \affiliation{\IKPJ}\affiliation{\JARA}
\author{A.~Szczurek}    \affiliation{\IFJ}
\author{A.~T\"aschner}  \affiliation{\MS}
\author{A.~Trzci{\'n}ski}\affiliation{\ASWarsN}
\author{R.~Varma}       \affiliation{\IITB}
\author{M.~Wolke}       \affiliation{\IKPUU}
\author{A.~Wro{\'n}ska} \affiliation{\IPJ}
\author{P.~W\"ustner}   \affiliation{\ZELJ}
\author{A.~Yamamoto}    \affiliation{\KEK}
\author{J.~Zabierowski} \affiliation{\ASLodz}
\author{M.J.~Zieli{\'n}ski}\affiliation{\IPJ}
\author{A.~Zink}        \affiliation{\Erl}
\author{J.~Z{\l}oma{\'n}czuk}\affiliation{\IKPUU}
\author{P.~{\.Z}upra{\'n}ski}\affiliation{\ASWarsN}
\author{M.~{\.Z}urek}   \affiliation{\IKPJ}

\newcommand*{\Mainz}{Institut f\"ur Kernphysik, Johannes 
 Gutenberg--Universit\"at Mainz, Johann--Joachim--Becher Weg~45, 55128 Mainz, 
 Germany}
\newcommand*{\EMAIL}{email address: mikhail.bashkanov@ed.ac.uk}
\newcommand*{\PGI}{Peter Gr\"unberg Institut, PGI--6 Elektronische 
 Eigenschaften, Forschungszentrum J\"ulich, 52425 J\"ulich, Germany}
\newcommand*{\DUS}{Institut f\"ur Laser-- und Plasmaphysik, Heinrich--Heine 
 Universit\"at D\"usseldorf, Universit\"atsstr.~1, 40225 D\"usseldorf, Germany}
\newcommand*{\Aachen}{III.~Physikalisches Institut~B, Physikzentrum, 
 RWTH Aachen, 52056 Aachen, Germany}

\collaboration{WASA-at-COSY Collaboration}\noaffiliation

%
%

\begin{abstract}
We report on a high-statistics measurement of the
most basic double pionic fusion reaction $\vec{n}p \to d\pi^0\pi^0$ over the
energy region of the $d^*(2380)$ resonance by use of a polarized deuteron beam
and observing the double fusion reaction in the quasifree scattering mode. The
measurements were performed with the WASA detector setup at COSY. The data
reveal substantial analyzing powers and confirm conclusions about the $d^*$
resonance obtained from unpolarized measurements. We also confirm the previous
unpolarized data obtained under complementary kinematic conditions.
\end{abstract}


\maketitle

\section{Introduction}

As has been pointed out previously by Harney \cite{harney}, finite vector
analyzing powers $A_y(\Theta)$ arise in reaction processes only, if at least
two different partial waves interfere. 
Hence in case of an isolated $s$-channel resonance, which is formed
by a single partial wave matching to spin and parity of the resonance, the
analyzing powers in the resonance region will be  vanishing small, if there is
no sizeable interfering background from other reaction processes.    

Recently, in the reaction $pn \to d\pi^0\pi^0$ a pronounced, narrow resonance
structure corresponding to a mass of 2.38 GeV and a width of about 70 MeV has
been observed in the total cross section near $\sqrt s \approx$ 
2.4 GeV ($T_p$ = 1.2 GeV) \cite{MB,prl2011,isofus}. Its quantum numbers have  been
determined to be $I(J^P) = 0(3^+)$ \cite{prl2011}. The $s$-channel character
of this resonance has been established recently by polarized $\vec{n}p$
scattering. Inclusion of these new data into the SAID partial-wave analysis
produces a pole in the coupled $^3D_3$-$^3G_3$ partial-waves at ($2380\pm10 -
i 40\pm5$) MeV \cite{prl2014,pnfull}. Since then this resonance is denoted by
$d^*(2380)$.   


The $d\pi^0\pi^0$ channel is the $d^*$ decay channel with the smallest amount
of background from other reaction processes \cite{prl2011,
  prl2014,pnfull,isofus,pp0-,np00,np+-,exa,hades}. Nevertheless it has sizeable contributions from
$t$-channel $N(1440)$ and $\Delta\Delta$ excitations. Both of them are very
well known from the study of $pp$-induced two-pion production
\cite{isofus,oset,IHEP,JJ,WB,JP,ae,Roper,iso,deldel,nnpipi,tt,FK}. 


Hence, due to the finite background amplitudes we may expect sizeable
analyzing powers $A_y$ in the region of the $d^*$ resonance. Also, they are
expected to increase with increasing energy due to the increasing contribution
of higher partial waves. Since $A_y$ is composed only of
interference terms of partial waves, it is sensitive to even small
partial-wave contributions and therefore qualifies as a sensitive
spectroscopic tool for the investigation of the $d^*$ resonance region.  

\section{Exclusive Measurements at WASA}

In order to investigate this issue in a comprehensive way we measured the 
basic {\it isoscalar} double-pionic fusion process $\vec{n}p \to d\pi^0\pi^0$
exclusively and kinematically complete. 

The experiment was carried out with the WASA detector setup \cite{barg,wasa}
at COSY  via the reaction $\vec{d}p \to d\pi^0\pi^0 + p_{spectator}$ using a
polarized deuteron beam at the lab energy $T_d$ = 2.27 GeV. Since due to Fermi
motion of the nucleons in the beam deuteron the quasifree reaction proceeds
via a range of effective collision energies, we cover the energy region 2.30
GeV $<\sqrt{s} <$ 2.47 GeV.
 
The emerging deuterons as well as the fast, quasifree scattered spectator
protons were detected in the forward detector of WASA and
identified by the $\Delta$E-E technique. Gammas from the $\pi^0$ decay were
detected in the central detector. 
 
That way the full four-momenta were determined for all particles of an
event. Since the reaction was  measured kinematically overdetermined,
kinematic fits with 6 overconstraints could be performed for each event.    
From the full kinematic information available for each event also the
relevant total energy in the $np$ system could be reconstructed for each event
individually.

By just having a different trigger these measurements have been obtained in
parallel to the ones for $np$ elastic scattering \cite{prl2014,pnfull}. The
trigger used for the detection of the $d\pi^0\pi^0$ events required at least
one hit in the forward detector and three neutral hits in the central detector.

For details of the experiment, in particular also with respect to the
determination of the beam polarization, checks for quasifree scattering and
the procedure for deriving $A_y$ from the data, see Ref. \cite{pnfull}.

For convenience the absolute normalization of the cross section data has been
obtained just by relative normalization to the datum of the total cross
section at $\sqrt s$ = 2.38 GeV published in Ref. \cite{isofus}.

\section{Results}

\subsection{Analyzing powers}


\begin{figure}
\centering
\includegraphics[width=0.88\columnwidth]{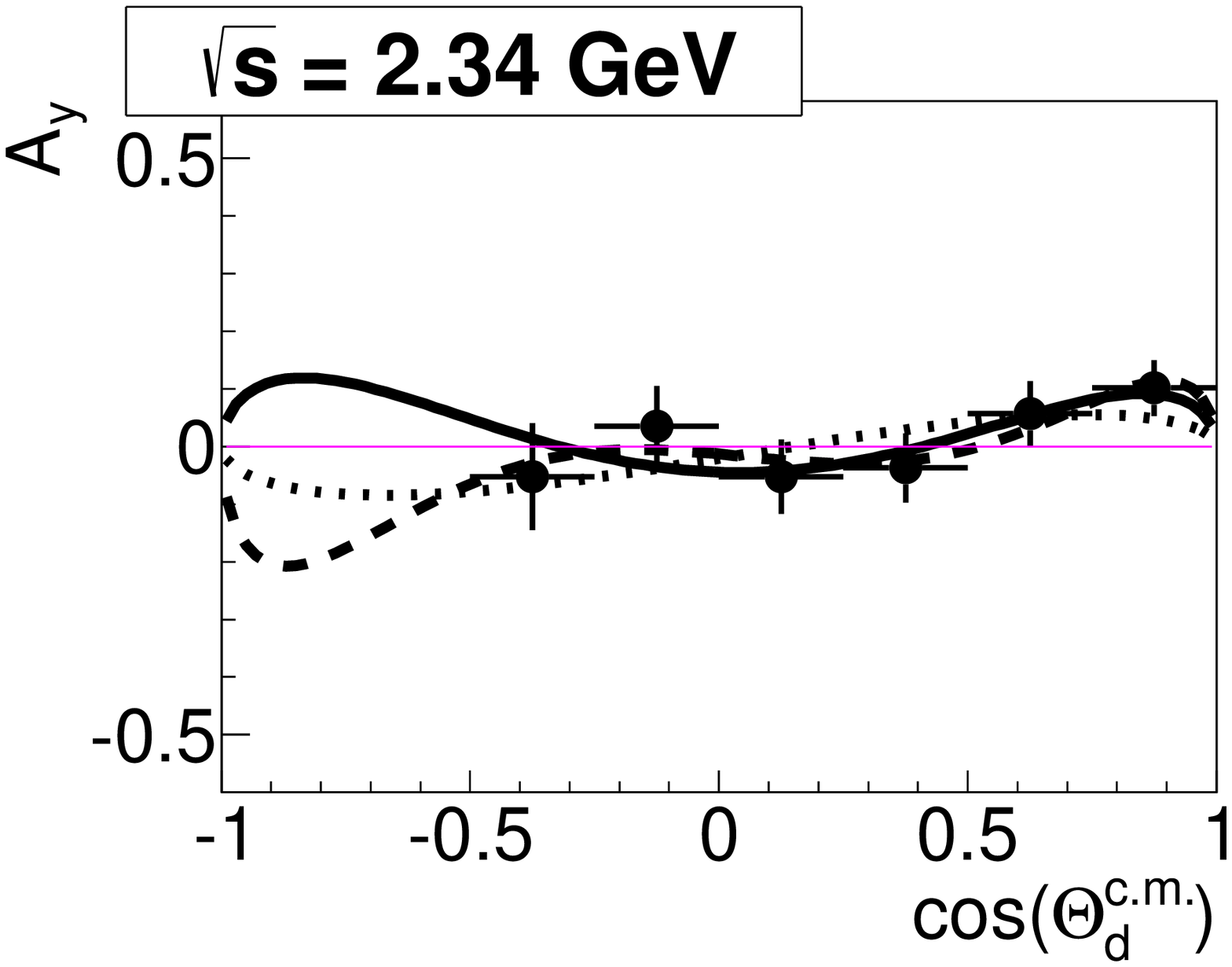}
\includegraphics[width=0.88\columnwidth]{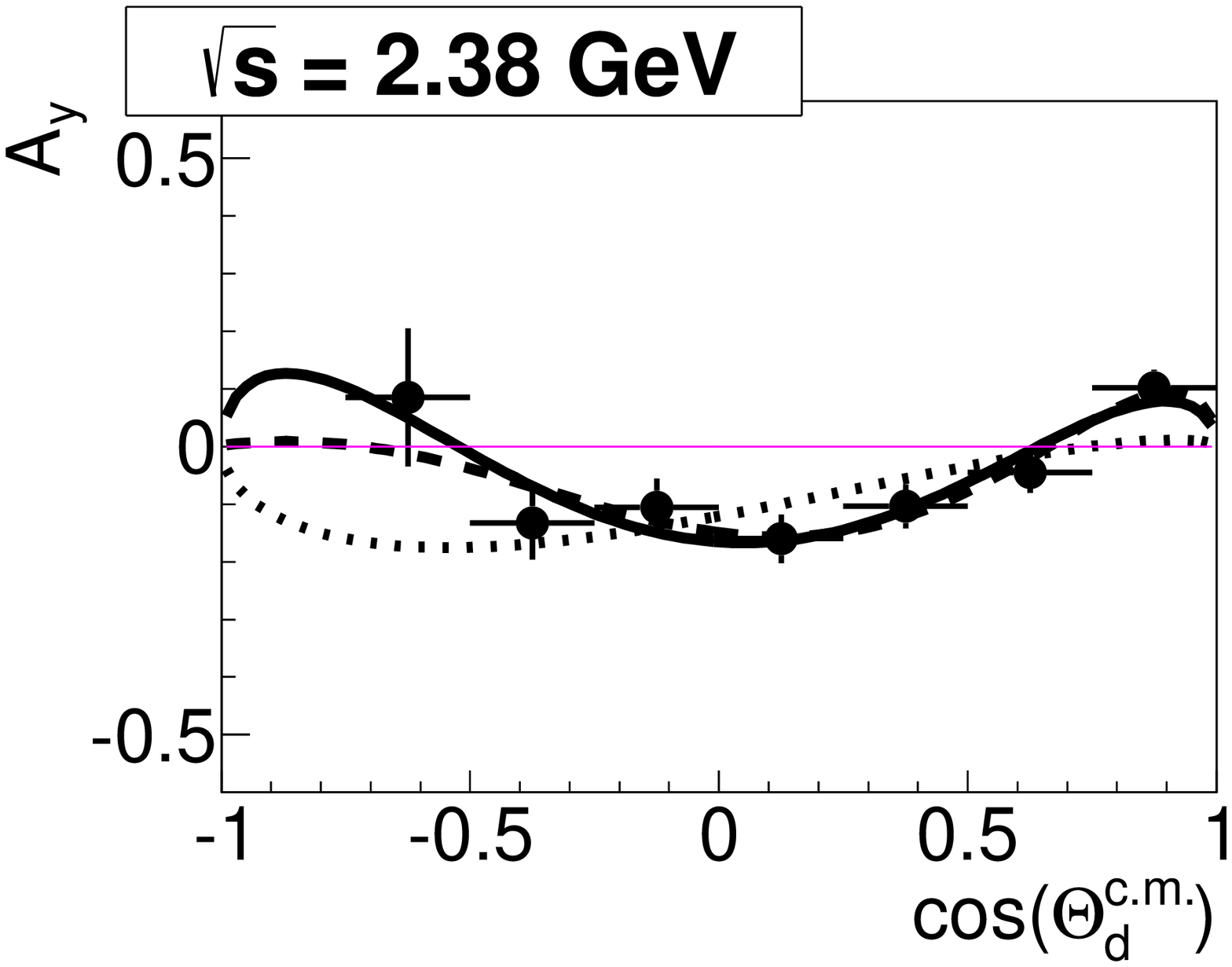}
\includegraphics[width=0.88\columnwidth]{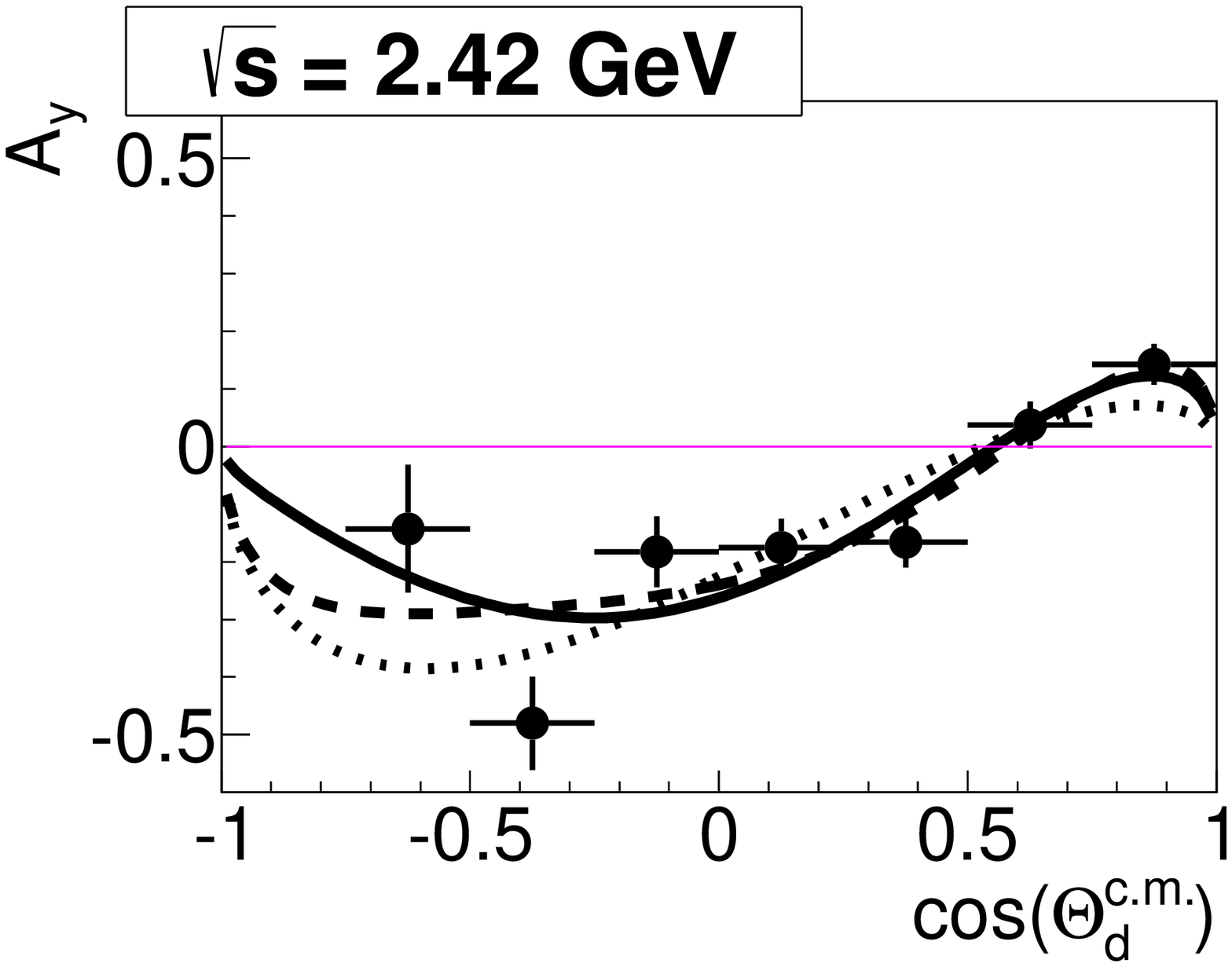}
\caption{\small   
Analyzing power in dependence of the deuteron scattering angle in the cm
system for the three energy bins centered at $\sqrt s$ = 2.34 GeV (top),
2.38 GeV (middle) and 2.42 GeV (bottom). The solid circles denote the
experimental results of this work. The dotted lines give a 2-parameter fit to
the data by use of eq.~(1). The solid lines show the fit results, if a $sin (3
\Theta^{cm})$ term is added and the dashed lines a fit, if also a $sin (4
\Theta^{cm})$ term is included, see eq. (2).  
}
\label{fig1}
\end{figure}


\begin{figure} 
\centering
\includegraphics[width=0.94\columnwidth]{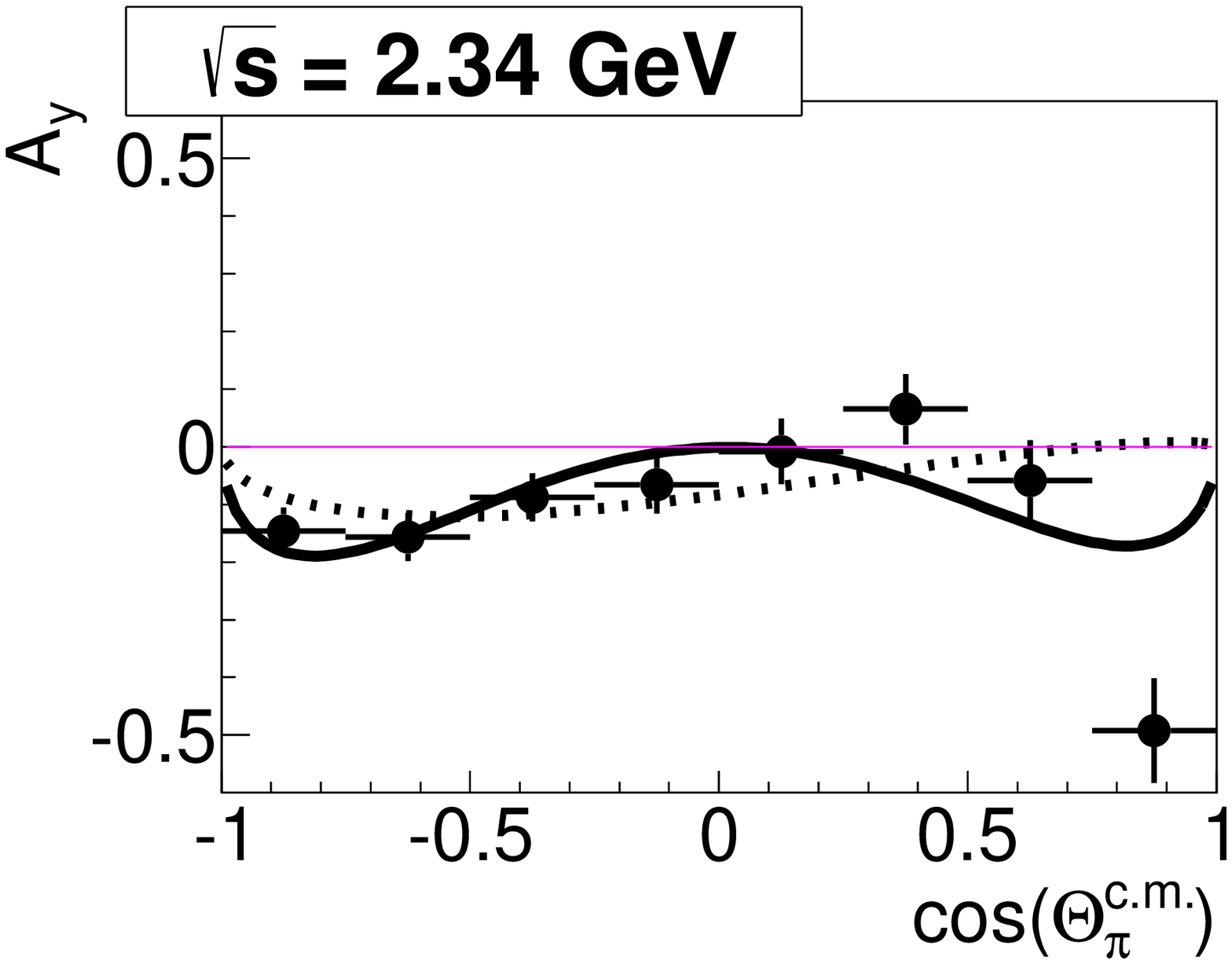}
\includegraphics[width=0.94\columnwidth]{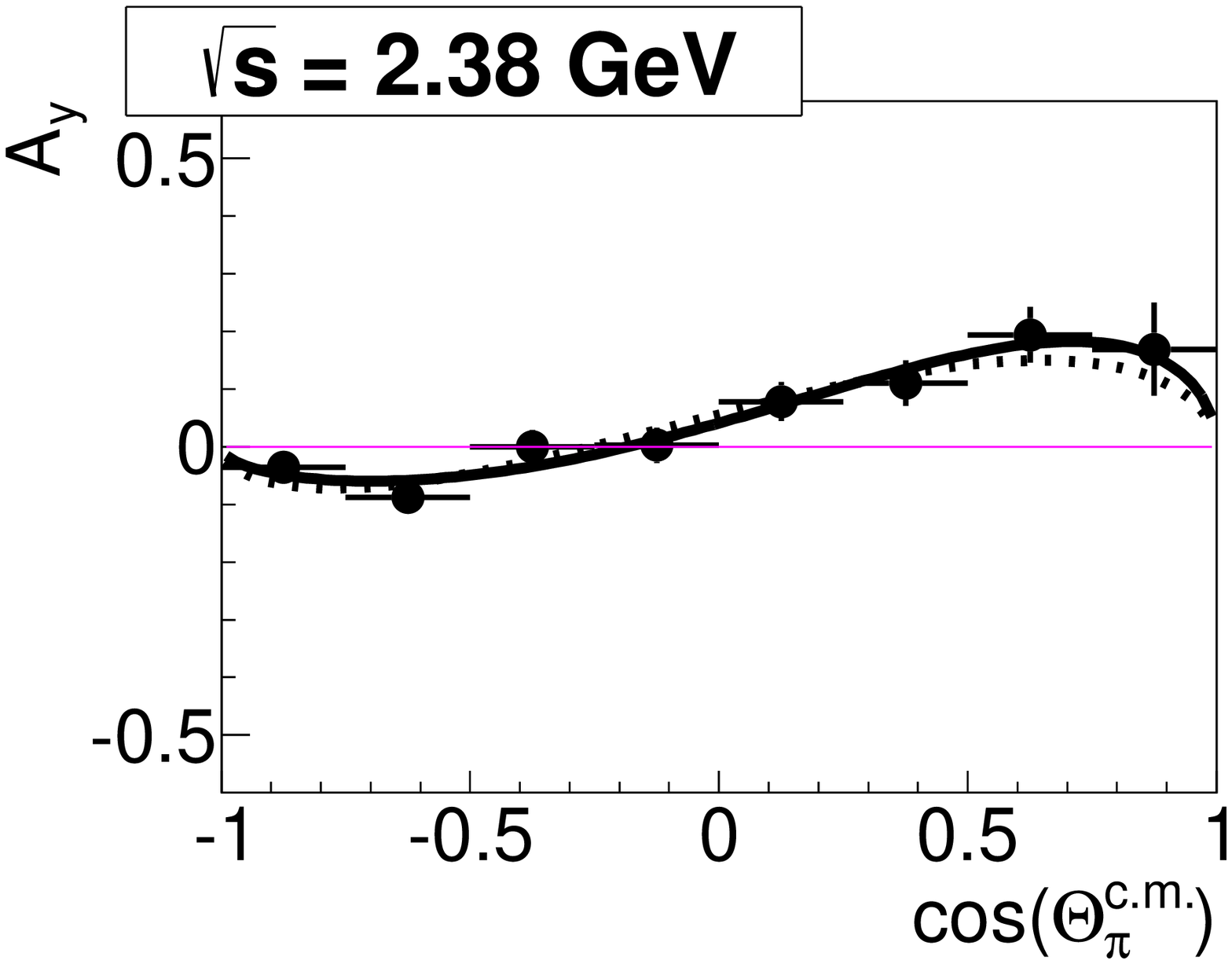}
\includegraphics[width=0.94\columnwidth]{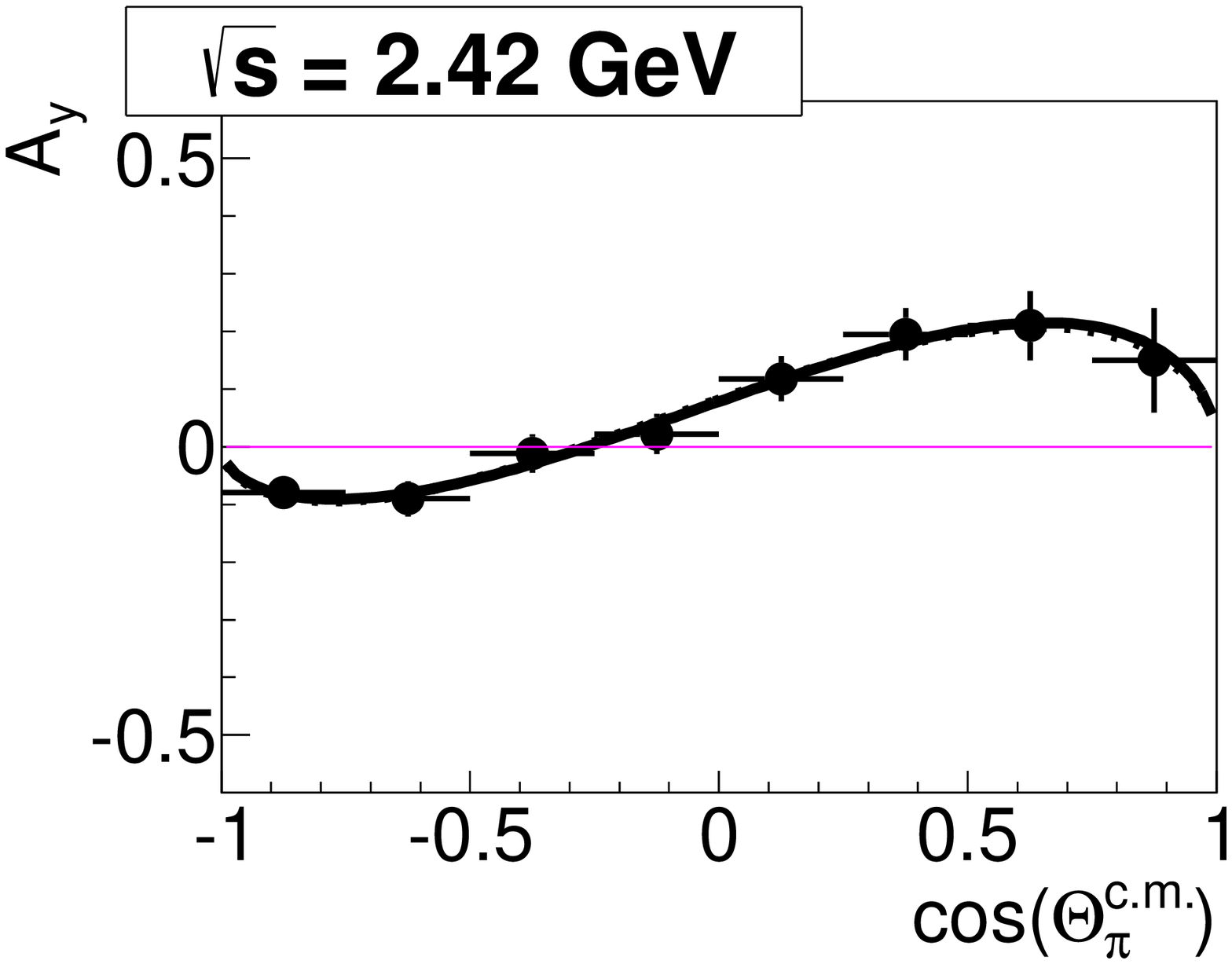}
\caption{\small   
The same as Fig.~1, but for the $\pi^0$ scattering angle in the cm
system. Fits are shown for the 2- and 3-parameter options.
}
\label{fig1}
\end{figure}


\begin{figure}
\centering
\includegraphics[width=0.94\columnwidth]{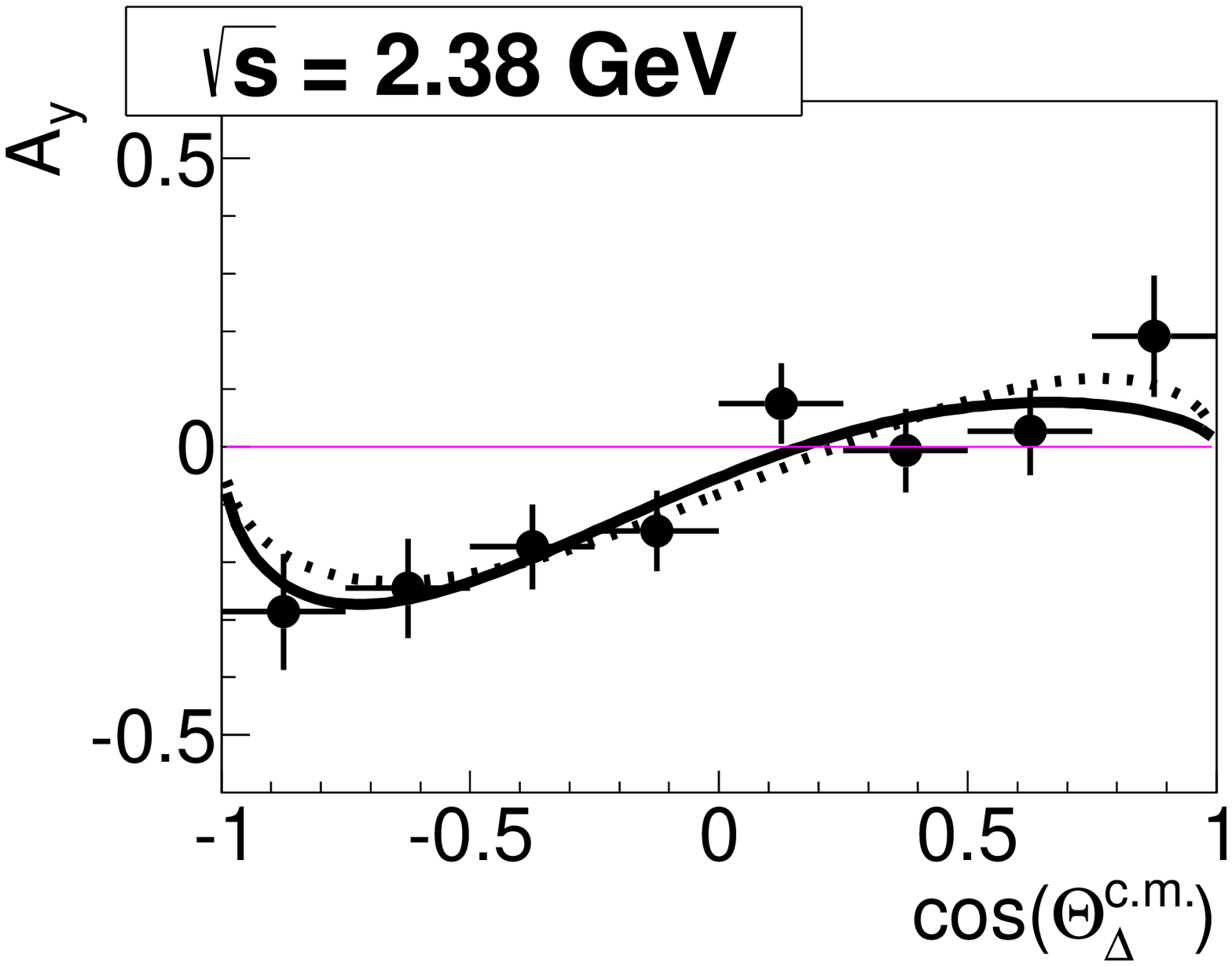}
\includegraphics[width=0.94\columnwidth]{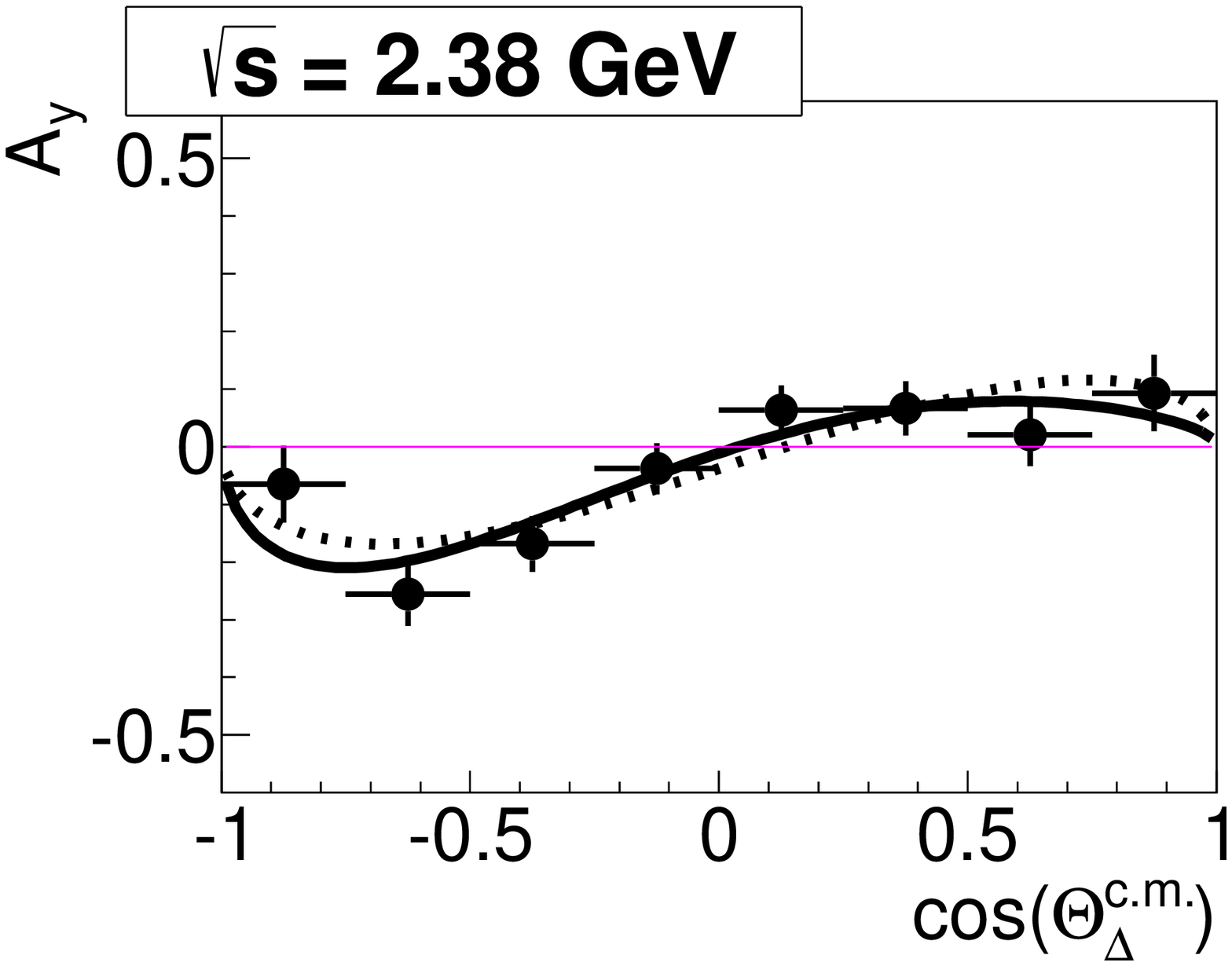}
\includegraphics[width=0.94\columnwidth]{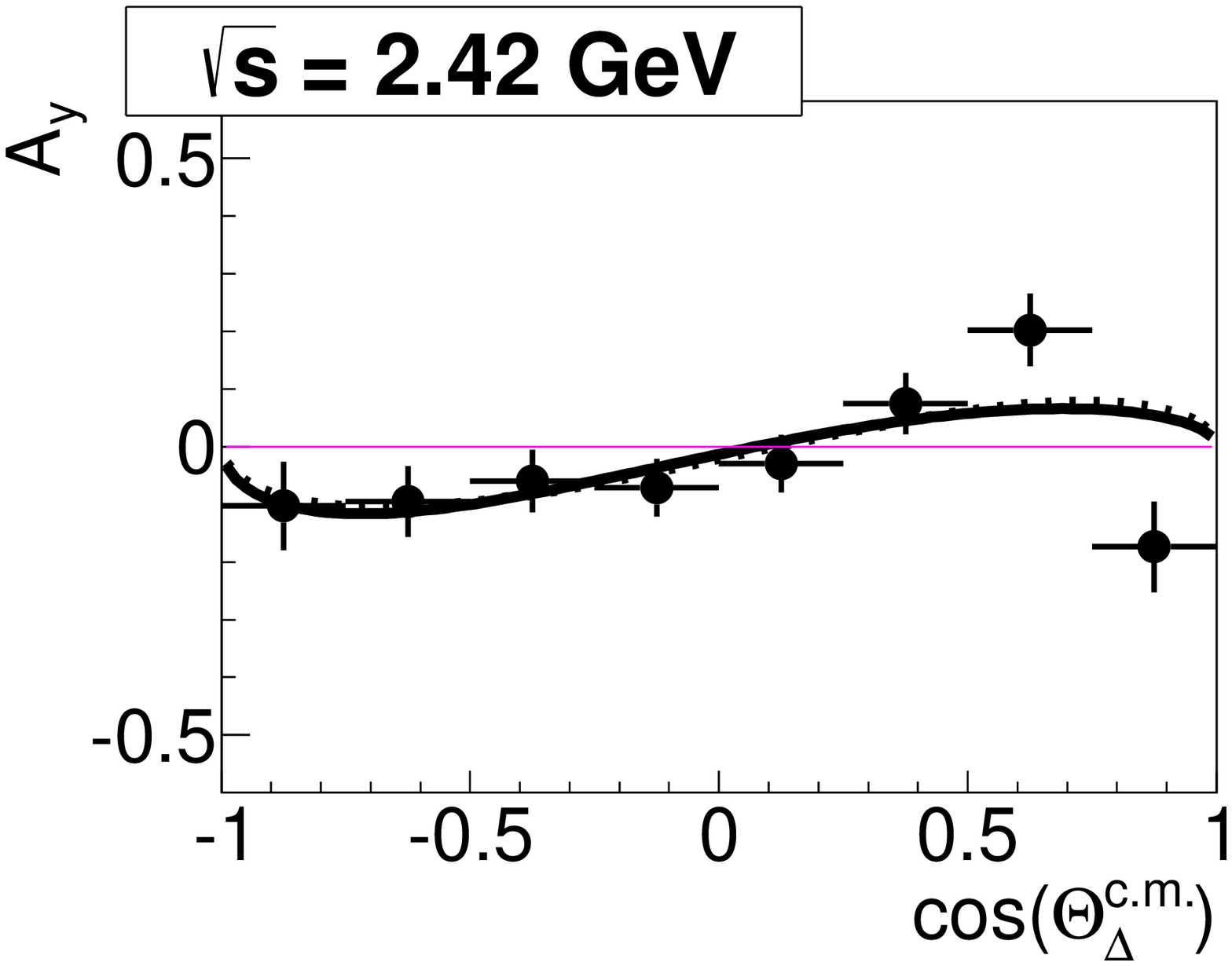}
\caption{\small   
The same as Fig.~2, but for the $\Delta$ scattering angle in the cm system.
}
\label{fig1}
\end{figure}

The analyzing powers $A_y$ extracted from this experiment are shown in Figs.~1
- 3  in dependence of the center-of-mass (c.m.) scattering angles
$\Theta_d^{c.m.}$, $\Theta_{\pi^0}^{c.m.}$ and  $\Theta_{\Delta}^{c.m.}$ of
emitted deuteron, $\pi^0$ and $\Delta$ particles, respectively. The
intermediate $\Delta$ from the process $d^* \to \Delta^+\Delta^0 \to
d\pi^0\pi^0$ has been reconstructed from the 4-momenta of its decay products
$\pi^0$ and nucleon -- the latter by taking half the deuteron momentum,
thereby neglecting the small correction due to Fermi motion of the nucleons
inside the deuteron. Since the Dalitz plot displayed in Fig.~4 of
Ref. \cite{prl2011} exhibits a $\Delta$ excitation band sitting upon no
substantial background, no cut on the $\Delta$ mass appears to be necessary.

The data have been binned into three energy bins as displayed in Figs.~1 - 3:
$\sqrt s$~=~2.30~-~2.35~GeV with center of gravity at 2.34~GeV , $\sqrt s$~=~
2.36~-~2.40~GeV with centroid at 2.38~GeV and $\sqrt s$~=~2.41~-~2.47~GeV
centered at 2.42~GeV.  The middle one corresponds to the
maximum cross section of the $d^*$ resonance, whereas the other two roughly
correspond to its half maximum. At the lowest energy bin the analyzing power
in dependence of the deuteron scattering angle is still small. However,
substantial $A_y$ values are obtained at the two higher energy bins.

In the following the description of the data is based on the formalism outlined
in Ref. \cite{CH}. Based on that work $A_y$ angular dependencies have been
derived in Ref.~\cite{ae}, which can 
be theoretically expected in $pp$ induced, {\it i.e.} purely isovector
two-pion production, if there are only relative $s$- and $p$- waves in the
final channel:    

\begin{equation}
A_y(\Theta^{c.m.}) = a~sin(\Theta^{c.m.}) + b~sin (2 \Theta^{c.m.})
\end{equation}
with the parameters a and  b to be adjusted to the data.

For the $pn \to d\pi^0\pi^0$ reaction the situation changes insofar as we
deal here with a purely isoscalar channel. In addition
$d$-waves have to be included, in order to allow the formation of
$d^*(2380)$. For simplicity we assume the $\pi\pi$ system to be in
relative $s$-wave. At least for the resonance formation this is well justified
\cite{prl2011}. Applying the formalism presented in Ref.~\cite{ae} to
this situation \cite{ch} we again end up with a formal description in terms of
$sin(j\Theta)$:   

\begin{equation}
A_y(\Theta^{c.m.}) = \sum p_{j-1}~sin(j\Theta^{c.m.})
\end{equation}

Due to the involvement of $d$-waves the sum runs now over 4 terms (j = 1,
... 4) from $sin(\Theta^{c.m.})$ until $sin(4\Theta^{c.m.})$. The weighting
parameters $p_0$ ... $p_3$ to be adjusted to the data have now the following
meaning: 

\begin{eqnarray}
p_0&= q a_1^* b + p_2~~~~~~~~~~~~~~~~~&(sp + pd)\\ \nonumber
p_1& = q^2 a_2^* r_1 + 4 p_3~~~~~~~~~~~~~ &(sd + dd)\\ \nonumber
p_2& = q^3 b r_1^*~~~~~~~~~~~~~~~~~~~~~~&(pd) \\ \nonumber
p_3& = q^4 r_1^* r_2~~~~~~~~~~~~~~~~~~~~&(dd) \\ \nonumber
\end {eqnarray}

Here $q$ denotes the momentum of the $\pi\pi$ system relative to the
deuteron and the strength parameters $a_1, a_2, b, r_1$ and $r_2$ stand for
the transitions:

\begin{eqnarray}
a_1: {^3S_1} \to {^3S_1} s~~~~~~~~~~~~~~~(s)\\ \nonumber
a_2: {^3D_1} \to {^3S_1} s~~~~~~~~~~~~~~~(s)\\ \nonumber
b  : {^1P_1} \to {^3S_1} p~~~~~~~~~~~~~~~(p)\\ \nonumber
r_1: {^3D_3} \to {^3S_1} d~~~~~~~~~~~~~~~(d)\\ \nonumber
r_2: {^3G_3} \to {^3S_1} d,~~~~~~~~~~~~~~(d)
\end{eqnarray}

where on the left-hand side the $pn$ partial wave in the entrance channel is
given by its spectroscopic nomenclature. The right-hand side denotes the partial
wave of the deuteron together with its angular momentum relative to the
$\pi\pi$ system. The interference of these partial waves, which are
abbreviated by $s$, $p$ and $d$, is indicated in brackets at the right-hand
side of eq. (3). Note that in the entrance channel $^3S_1$ and $^3D_1$ as well
as $^3D_3$ and $^3G_3$ are coupled partial waves. In principle, also the $^3S_1$
- $^3D_1$ coupled waves contribute to the $^3S_1 d$ configurations. However,
for simplicity we omit this contribution, since it is expected to be small
compared to the contribution of the $d^*$ resonance.  

In order to see how many terms in the expansion (3) are needed by the data, we
performed fits with 2, 3 and 4 terms as given in Tables 1 - 3 and shown in
Figs. 1 - 3 by the dotted, solid and dashed lines, respectively. 

For the
analyzing power in dependence of the deuteron scattering angle the latter two
are very close together in the angular regions, which are well covered by
data. This means that a 3-parameter fit is already appropriate for a proper
description of the data. For the lowest energy, where the data are very close
to zero throughout the measured angular range, already the 2-parameter fit
is sufficient with providng a $\chi^2$ per degree of freedom (ndf) of
unity. The fact that already a 3-parameter fit is sufficient for an
appropriate description of the data in the resonance region is in
accordance with the new SAID solution, which exhibits the $d^*$ pole
predominantly in the $^3D_3$ wave and only very weakly in $^3G_3$. Hence $r_2$
got to be small and $p_3$ negligible compared to $p_2$. In fact, the resonance
term $p_2$ is highly demanded by the data, as the comparison between dashed
and dotted curves demonstrates. Since $p_2$ enters also in $p_0$, the latter
is also requested by the fit, whereas $p_1$ turns out compatible with zero
within uncertainties at resonance. Therefore, the leading
contribution to the analyzing 
power of the $\Theta_d^{c.m.}$ angular distribution turns out to be the
interference of the resonant $d$ wave with the non-resonant $p$ wave.

\begin{table}
\caption{Results of the fits to the analyzing power data in dependence of the
  deuteron scattering angle by use of eq. (2) with two (2p), three (3p) and
  four (4p) terms.}  
\begin{tabular}{llllllll}
\\ 
\hline

$\sqrt s$&fit&$p_0$&$p_1$&$p_2$&$p_3$&$\chi^2/ndf$\\ 
(GeV)& \\

\hline

2.34&4p&-.04(10)&.11(15)&-.03(13)&.07(8)&0.9/2\\ 
    &3p& .04(5)&-.01(7)&.08(6)&         &1.8/3\\ 
    &2p&-.02(4)&-.07(4)&      &         &4.1/4\\
2.38&4p&-.07(4)& .01(6)&.08(5)&.03(4)&2.8/3\\
    &3p&-.04(3)&-.03(4)&.12(3)&      &3.6/4\\
    &2p&-.12(2)&-.08(3)&      &      &19/5\\
2.42&4p&-.20(4)& .18(6)&.04(6)&.04(4)&11/3\\ 
    &3p&-.17(4)& .14(4)&.09(3)&      &12/4\\  
    &2p&-.22(3)& .21(3)&      &      &19/5\\
\hline
 \end{tabular}\\
\end{table}

\begin{table}
\caption{Results of the fits to the analyzing power data in dependence of the
  $\pi^0$ scattering angle by use of eq. (2) with two (2p) and three (3p)
  terms.}   
\begin{tabular}{lllllll}
\\ 
\hline

$\sqrt s$&fit&$p_0$&$p_1$&$p_2$&$\chi^2/ndf$\\  
(GeV)& \\

\hline

2.34&3p&-.12(3)&.01(3)&-.12(3)&20/5\\
    &2p&-.08(2)&.06(3)&       &39/6\\ 
2.38&3p& .06(2)&.12(2)& .02(2)&3.6/5\\  
    &2p& .06(2)&.11(2)&       &5.1/6\\ 
2.42&3p& .08(2)&.15(2)& .00(2)&0.9/5\\ 
    &2p& .08(2)&.15(2)&       &0.9/6\\  
\hline
 \end{tabular}\\
\end{table}

\begin{table}
\caption{Results of the fits to the analyzing power data in dependence of the
  $\Delta$ scattering angle by use of eq. (2) with two (2p) and three (3p)
  terms.}   
\begin{tabular}{lllllll}
\\ 
\hline

$\sqrt s$&fit&$p_0$&$p_1$&$p_2$&$\chi^2/ndf$\\  
(GeV)& \\

\hline

2.34&3p&-.10(4)&.17(4)&-.04(4)&5.0/5\\
    &2p&-.08(3)&.17(4)&       &6.0/6\\ 
2.38&3p&.-05(2)&.14(3)&.-04(3)&7.6/5\\ 
    &2p&-.04(2)&.14(3)&       &9.9/6\\ 
2.42&3p&.-02(3)&.09(3)&-.01(3)&15/5\\  
    &2p&-.02(2)&.09(3)&       &15/6\\ 
\hline
 \end{tabular}\\
\end{table}

The $q$-dependence of the parameters makes it plausible that the analyzing
power is smallest at the lowest energy $\sqrt s$ = 2.34 GeV and tends to level
off as soon as the resonance maximum is reached. At 2.42 GeV the resonance
amplitude is already substantially reduced, however, the $q$-dependence in
$p_0$ and $p_2$ counteracts this reduction.

For the $\Theta_{\pi^0}^{c.m.}$ dependence of the analyzing power we may stick
with the same ansatz eq. (2), but need to reinterpret the transitions (4) with
respect to the partition $d\pi^0$~-~$\pi^0$. With still having the
$\pi^0\pi^0$ system coupled to zero, this means that the transitions $(s)$ and
$(d)$ both represent configurations, where the $\pi^0$ is in relative $p$ wave
to the $d\pi^0$ system, {\it i.e.} contain also resonance contributions. If we
forget the somewhat erratic data point at small 
angles at $\sqrt s$ = 2.34 GeV, then we observe an approximately constant
pattern over the energy region of interest, which can be described
sufficiently well by already the first two terms in the expansion eq. (2).

Finally, for the $\Delta\Delta$ partition we expect relative $s$-waves
independent of whether this partition originates from $d^*$ or conventional
$t$-channel excitation, since the considered energies are still below the
nominal mass of two $\Delta$ excitations. The observed $A_y$ distributions are
similar to those for the $d\pi^0$~-~$\pi^0$ partition and hence
characterized dominantly by the $p_1$ contribution. 

\subsection{Cross sections}

By using both the unpolarized and polarized runs of this experiment we may
extract also (unpolarized) differential and total cross sections. This is
valuable, since we 
used in this experiment the quasifree $pn$ collision in reversed kinematics
covering thus the lab system phase space complementary to what has been
obtained in regular kinematics used previously \cite{prl2011}. 

Fig.~4 shows the $\Theta_d^*$ angular dependence of the (unpolarized )
differential cross section over the energy region $\sqrt s$ = 2.33 - 2.43 GeV
binned into five intervals. The data plotted by the open circles have been
obtained in a previous experiment \cite{prl2011} by use of a proton beam
hitting a deuterium target in quasi-free kinematics. Due to the experimental
conditions only the deuteron back-angles could be measured in good
quality. Now, with a deuteron beam impinging on a hydrogen target the phase
space in the lab system is populated in a complementary way and we may deduce
the cross sections preferably at forward angles (solid circles). Note that in
Fig.~4 the data are plotted at angles mirrored to the way plotted in Fig.~5 of
Ref. \cite{prl2011}. Here, in this work, the angles are defined relative to the
direction of the initial neutron, whereas in Ref. \cite{prl2011} they have been
defined relative to the direction of the initial proton.

Since we deal here with a purely isoscalar reaction, the unpolarized angular
distributions 
have to be symmetric about $90^\circ$ in the cms (Barshay-Temmer theorem
\cite{BT}). The data are in very good agreement with this requirement. To
underline this, we show in Fig. 4 fits with an expansion into Legendre
polynomials of order 0, 2, 4 and 6, {\it i.e.} including $d$-waves between d and
$\pi^0\pi^0$ systems and allowing total angular momenta up to $J_{max}$ = 3:

\begin{equation}
\sigma(\Theta^{c.m.}) = \sum_{n=0}^{J_{max}} a_{2n}~P_{2n}(\Theta^{c.m.}),
\end{equation}


where the coefficients $a_{2n}$ denote the fit parameters.

In addition to the symmetry about $90^\circ$ Fig.~4 demonstrates that the
anisotropy is largest around the maximum of the $d^*$ resonance flattening off
below and above. The fact that the angular distribution tends to flatten out
towards lower energies is not unexpected, since close to threshold we expect
contributions only from the lowest partial waves. The fact that the angular
distribution tends to be flatter also at the high energy end of the
investigated energy region, is not as trivial. It supports the fact that the
high spin $J$ = 3 of the $d^*$ resonance requires a unusually large
anisotropy of the angular distribution, which is larger than obtained in the
conventional $t$-channel $\Delta\Delta$ process, which gets the dominant
mechanism at higher energies and where the $\Delta\Delta$ system may also be in
lower angular momentum configurations.

\begin{figure} 
\begin{center}
\includegraphics[width=0.49\columnwidth]{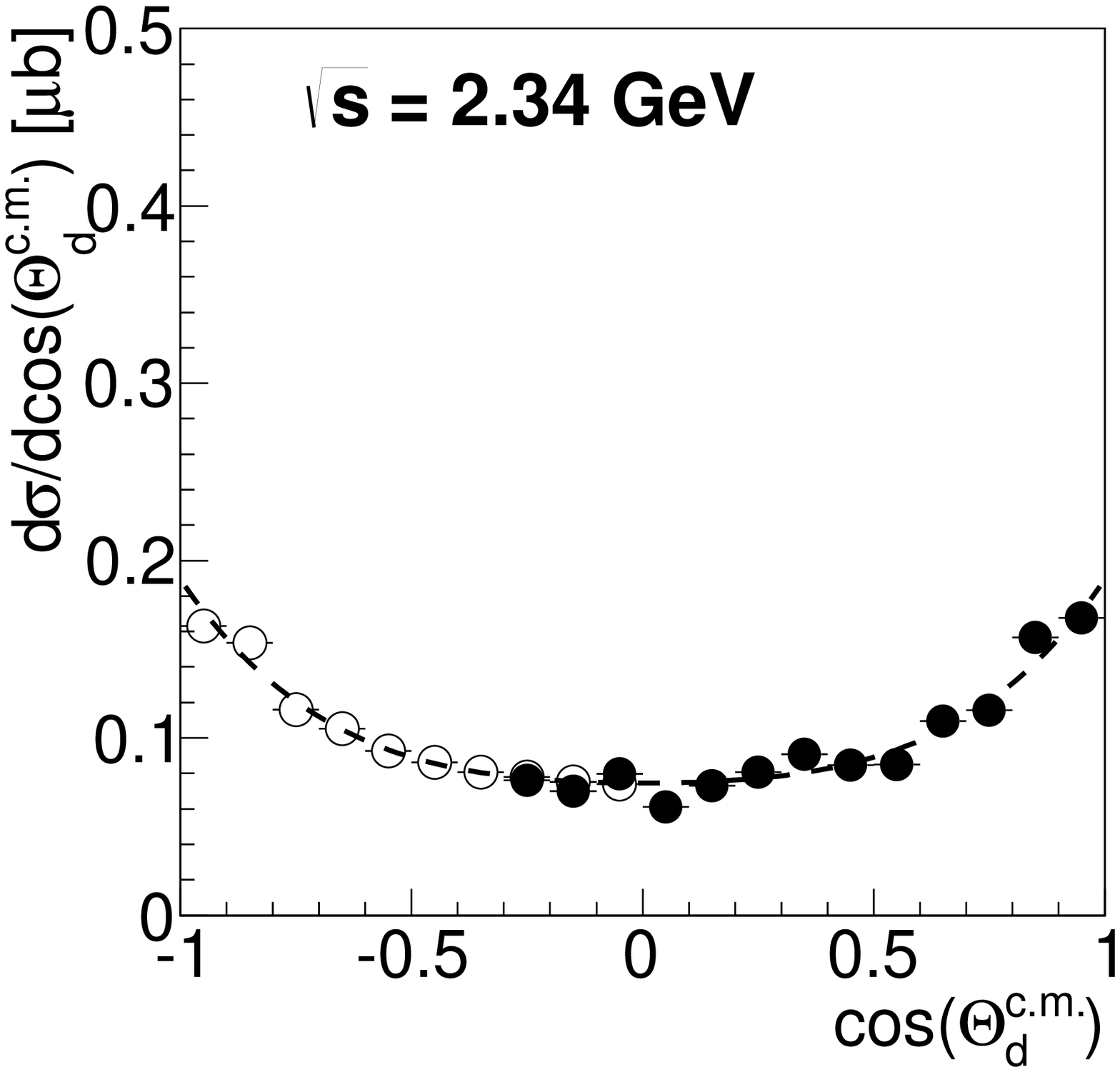}
\includegraphics[width=0.49\columnwidth]{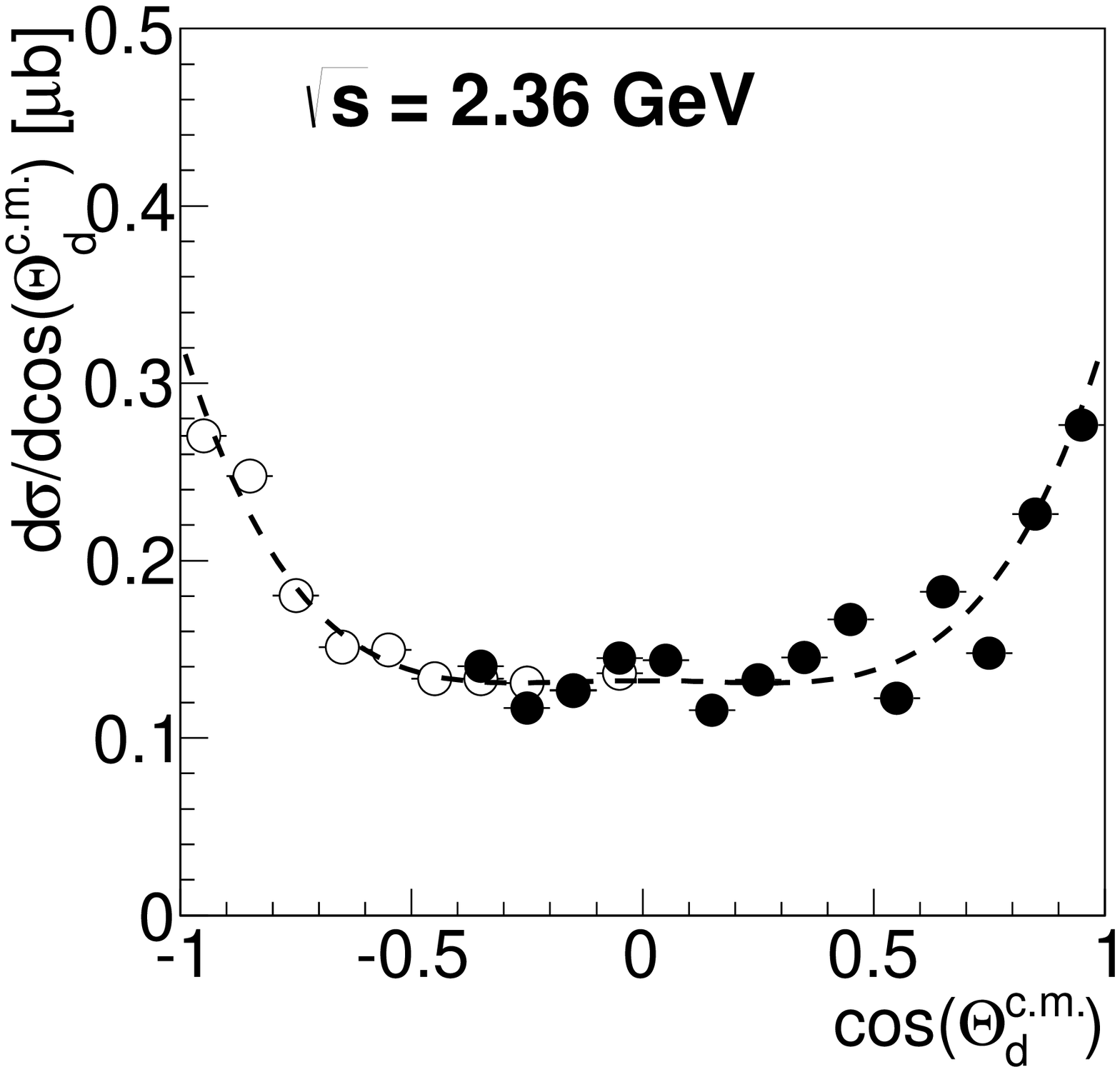}
\includegraphics[width=0.49\columnwidth]{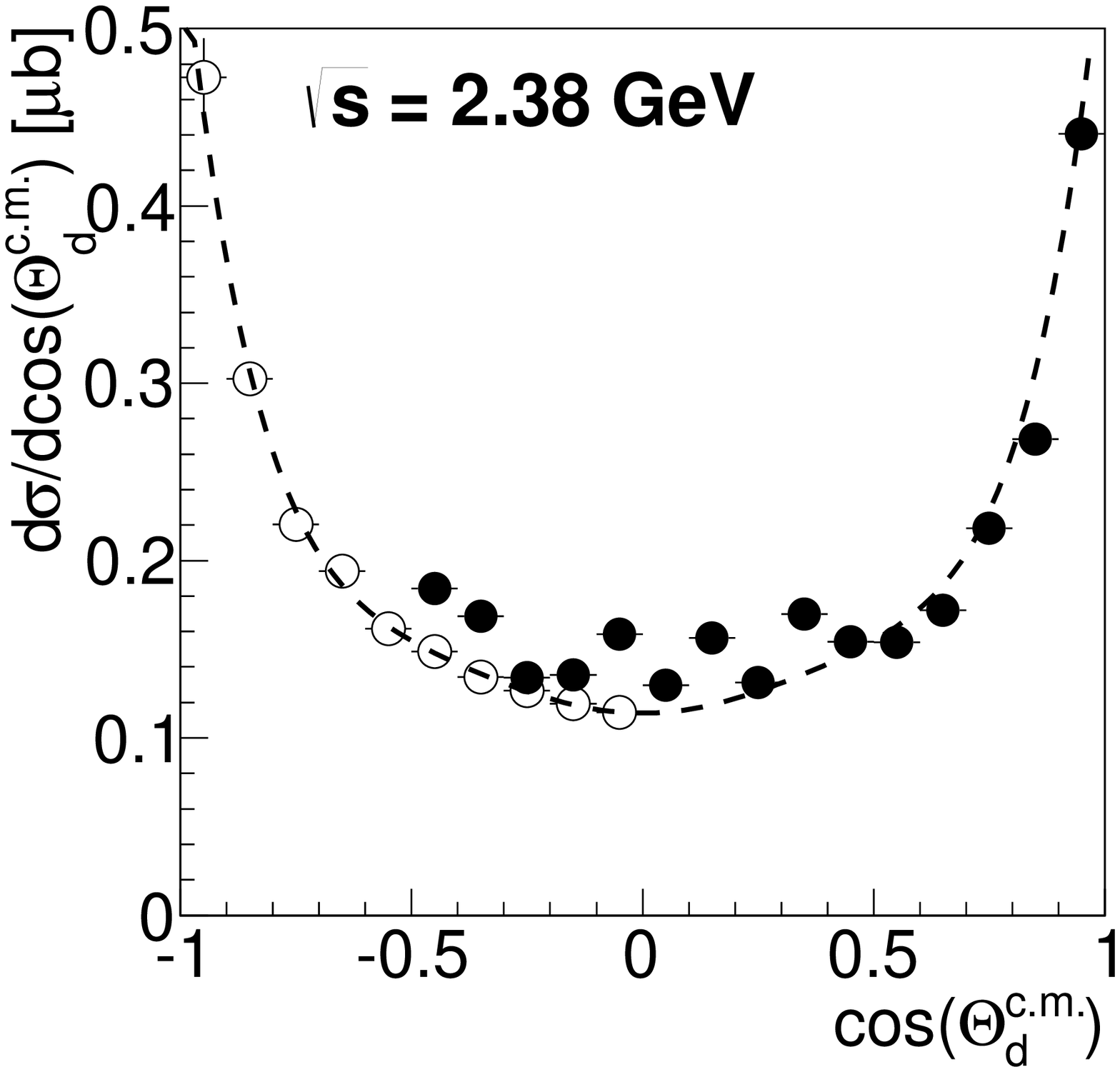}
\includegraphics[width=0.49\columnwidth]{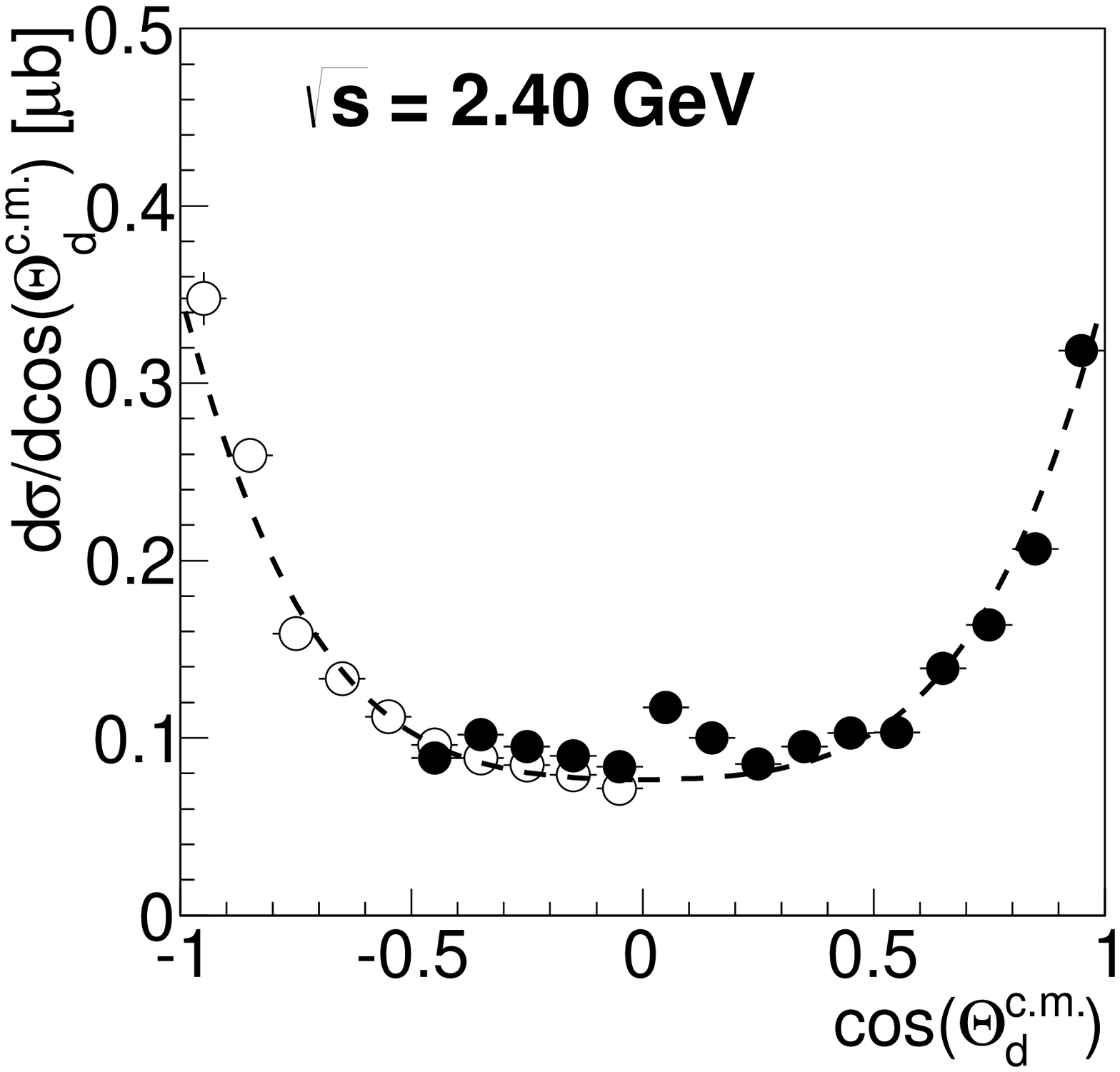}
\includegraphics[width=0.49\columnwidth]{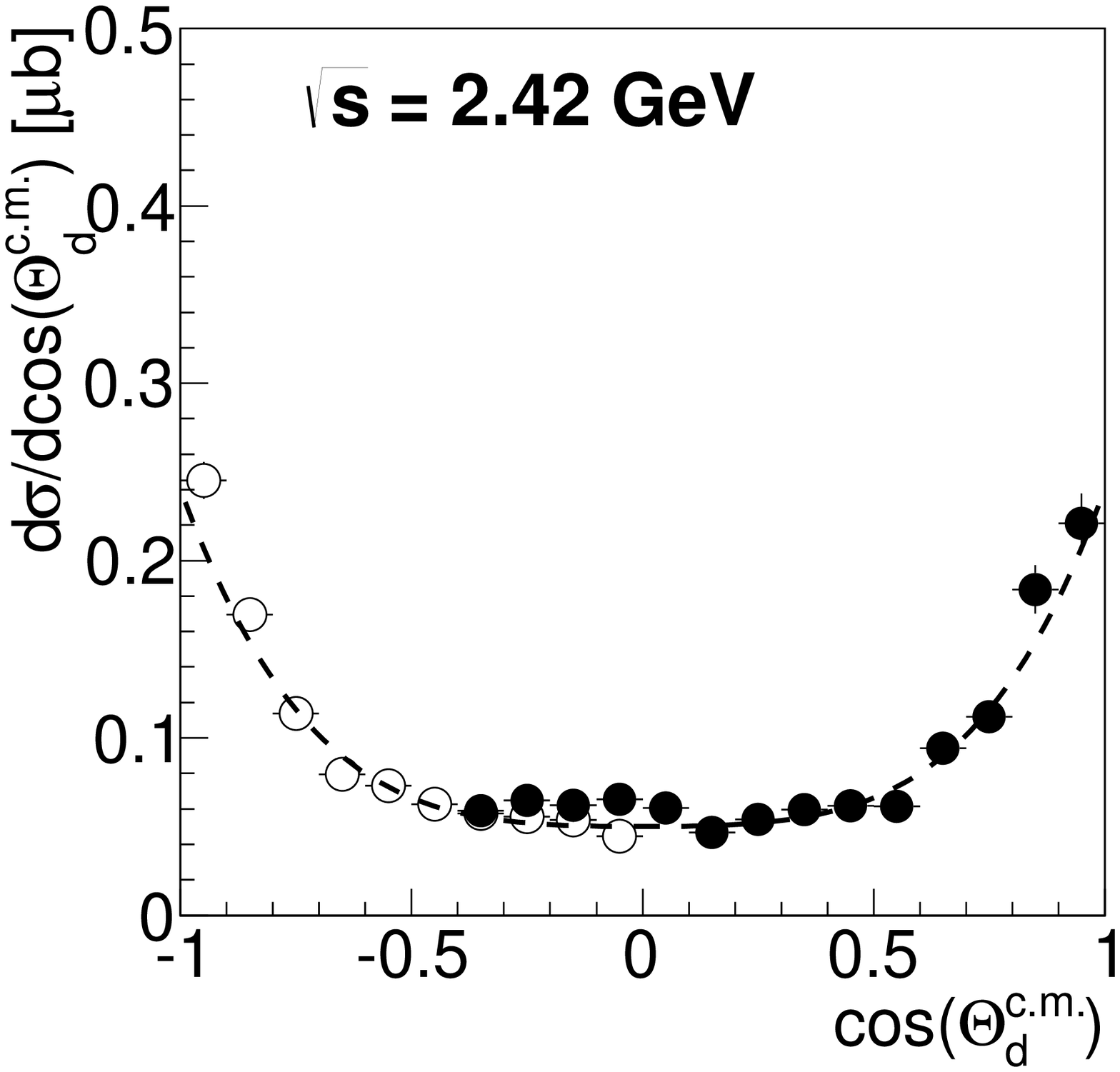}
\caption{\small Deuteron angular distributions across the energy region $\sqrt
  s$ = 2.33 - 2.43 GeV binned into five intervals. Open circles denote
  previous results \cite{prl2011}, filled circles this work. The dashed curves
  give Legendre fits with $L_{max} \leq$ 6.  
}
\end{center}
\end{figure}

Finally we display in Fig.~5 the energy dependence of the total cross section as
obtained with three independent measurements under different experimental
conditions: 
\begin{itemize}
\item  $pn$ collisions under usual quasifree kinematics with unpolarized beam
  and without magnetic field in the central part of the WASA detector at three
  beam energies (open circles \cite{prl2011}),
\item $pn$ collisions under usual quasifree kinematics with unpolarized beam,
  but with magnetic field in the central part of the WASA detector (open
  diamonds \cite{isofus}) and
\item $\vec{n}p$ collisions under reversed quasifree kinematics with polarized
  beam and without magnetic field in the  central part of the WASA detector
  (filled circles, this work).
\end{itemize}

The data of the first and third measurements have been normalized in absolute
height to the value obtained in the second measurement \cite{isofus} for
$\sqrt s$ = 2.38 GeV. Within uncertainties the data from all three
experiments agree to each other.

\begin{figure} 
\begin{center}
\includegraphics[width=0.99\columnwidth]{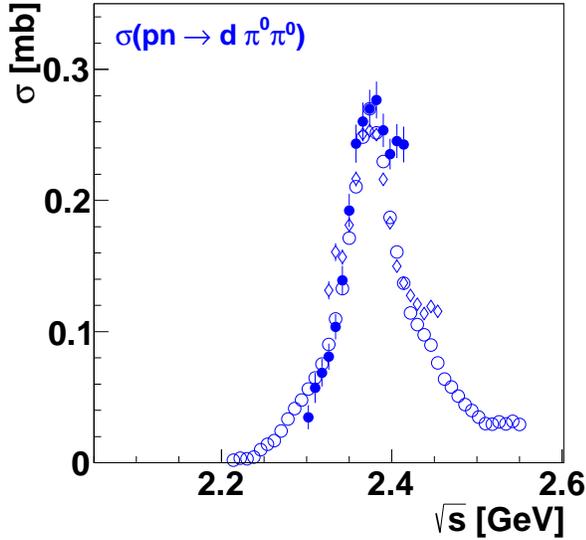}
\caption{\small Energy dependence of the total cross section measured under
  three different experimental conditions described in Refs. \cite{prl2011}(
  open circles), \cite{isofus} (open diamonds) and this work (filled circles),
  see text.  
}
\end{center}
\end{figure}

\section{Summary}
We have presented the first measurements of the $\vec{n}p \to d\pi^0\pi^0$
reaction 
with polarized beam using the quasifree $np$ collision process under reversed
kinematics. The deduced total cross sections are consistent with previous
results. The obtained deuteron angular distributions complement the previous
results. They clearly show that at resonance the anisotropy is larger than
outside. 

The measurements exhibit significant analyzing powers in dependence of
deuteron and pion angles, which can be understood as being due to the
interference of the $d^*$ resonance amplitude with background amplitudes. 

\section{Acknowledgments}

We acknowledge valuable discussions with Ch. Hanhart  
 on this issue. 
This work has been supported by Forschungszentrum J\"ulich
(COSY-FFE), DFG (CL 214/3-1), STFC (ST/L00478X/1), Foundation for Polish
Science through the MPD programme and by the Polish National Science Centre
through the Grants No.2011/01/B/ST2/00431, 2013/11/N/ST2/04152,
2011/03/B/ST2/01847.


\begin{thebibliography}{99}

\bibitem{harney} H. L. Harney, Phys. Lett. B {\bf 22}, 249 (1968).
\bibitem{MB} M. Bashkanov {\it et al.}, Phys. Rev. Lett. {\bf 102}, 052301 
  (2009).
\bibitem{prl2011} P. Adlarson {\it et al.}, Phys. Rev. Lett. {\bf 106}, 242302 
  (2011).
\bibitem{isofus} P. Adlarson \emph{et. al.}, Phys. Lett. B {\bf 721},
  229 (2013).   
\bibitem{prl2014} P. Adlarson {\it et al.}, Phys. Rev. Lett. {\bf 112}, 202301 
  (2014).
\bibitem{pnfull} P. Adlarson {\it et al.}, Phys. Rev. C {\bf 90}, 035204
  (2014).  
\bibitem{pp0-} P. Adlarson \emph{et. al.},  Phys. Rev. C {\bf 88}, 055208
  (2013).
\bibitem{np00} P. Adlarson \emph{et. al.},  Phys. Lett. B {\bf 743}, 325
  (2015).
\bibitem{np+-} H. Clement, M. Bashkanov and T. Skorodko,  Phys. Scr.T {\bf
    166},  014016 (2016).
\bibitem{exa} M. Bashkanov, H. Clement and T. Skorodko, Hyperfine Interact
  {\bf 234}, 57 (2015).
\bibitem{hades} G. Agakishiev  \emph{et. al.}, Phys. Lett. B {\bf 750}, 184
  (2015). 
\bibitem{oset} L. Alvarez-Ruso, E. Oset and E. Hernandez, Nucl. Phys. A {\bf
633}, 519 (1998).
\bibitem{IHEP} Xu Cao, Bing-Song Zou and Hu-Shan Xu, Phys. Rev. C {\bf 81},
  065201 (2010).
\bibitem{JJ} J. Johanson {\it et al.}, Nucl. Phys. A {\bf 712}, 75 (2002).
\bibitem{WB} W. Brodowski \emph{et. al.}, Phys. Rev. Lett. {\bf 88}, 192301
  (2002). 
\bibitem{JP} J. P\"atzold \emph{et. al.}, Phys. Rev. C {\bf 67}, 052202(R)
  (2003). 
\bibitem{ae}  S. Abd El-Bary \emph{et. al.}, Eur. Phys. J. A {\bf 37}, 267
  (2008). 
\bibitem{Roper} T. Skorodko \emph{et. al.}, Eur. Phys. J. A {\bf 35}, 317
  (2008).
\bibitem{iso} T. Skorodko \emph{et. al.}, Phys. Lett. B {\bf 679}, 30 (2009).
\bibitem{deldel} T. Skorodko \emph{et. al.}, Phys. Lett. B \textbf{695}, 115
  (2011). 
\bibitem{nnpipi} T. Skorodko \emph{et. al.}, Eur. Phys. J. A \textbf{47}, 108
  (2011).
\bibitem{tt} P. Adlarson {\it et al.}, Phys. Lett. B {\bf 706}, 256 (2011).
\bibitem{FK} F. Kren {\it et al.}, Phys. Lett. B {\bf 684}, 110 (2010) and B
  {\bf 702}, 312 (2011); arXiv:0910.0995v2 [nucl-ex]. 


\bibitem{wasa} H. H. Adam et al., arxiv: nucl-ex/0411038.
\bibitem{barg} Chr. Bargholtz {\it et al.}, Nucl. Instr. Meth {\bf A594}, 339
  (2008).
\bibitem{CH} Ch. Hanhart, Phys.\ Rept.\  {\bf 397}, 155 (2004).
\bibitem{ch} Ch. Hanhart, priv. comm.
\bibitem{BT} S. Barshay and G. M. Temmer, Phys. Rev. Lett. {\bf 12}, 728 (1964).
  
\end{thebibliography}
\end{document}